\documentclass[twocolumn,showpacs,aps,prb]{revtex4}

\usepackage{epsfig}

\newcommand{\vf}[1]{\mbox{\boldmath $#1$}}

\begin{document}

\title{Field-induced incommensurate order for the quasi-one-dimensional XXZ model in a magnetic field}

\author{Kouichi Okunishi and Takahumi Suzuki$^1$}

\affiliation{Department of Physics, Niigata University, Niigata 950-2181, Japan \\
$^1$Institute for Solid State Physics, University of Tokyo, Kashiwa, Chiba 277-8581, Japan}
\begin{abstract}
We investigate phase transitions of the quasi-one-dimensional $S=1/2$ XXZ model in a magnetic field, using the bosonization combined with the mean-field treatment of the inter-chain interaction.
We then find that the field induced incommensurate order is certainly realized in the low field region, while the transverse staggered order appears in the high field region.
On the basis of the result, we discuss the field-induced phase transition recently observed for BaCo$_2$V$_2$O$_8$. 
\end{abstract}

\maketitle

\section{introduction}

Field-induced phase transition in the quantum spin systems has been providing interesting physics such as magnon Bose Einstein condensation\cite{nikuni}.
Recently an exotic field induced phase transition was observed for  BaCo$_2$V$_2$O$_8$\cite{he}, which can be regarded as the quasi-one-dimensional(1D) $S=1/2$ XXZ antiferromagnt having the Ising-like anisotropy $\Delta\simeq 2$;  the magnetization and electron spin resonance (ESR) measurements above 1.8K show that BaCo$_2$V$_2$O$_8$ is basically described by the Bethe-ansatz-based theoretical analysis.\cite{kimura1} 
However, the specific heat measurements  up to 12T below 1.8K has revealed that the weak 3D couplings possibly trigger the exotic incommensurate(IC) order  in the low-field region.\cite{kimura2}
A peculiar point on this phase is that the ordering is different from the N\'eel type at the zero magnetic field and  the shape of the phase boundary in the $H$-$T$ plane is quite different from the usual field induced order in the coupled Haldane system\cite{honda}. 
This suggests that the Ising-like anisotropy in the quasi-1D system plays an essential role in the field induced IC order phase, behind which there is substantially important physics.

The 1D XXZ antiferromagnet is an exactly solved model playing the essential role to understand the critical quantum fluctuation and strong correlation effects.\cite{yang-yang,haldane,korepin} 
Although the Ising-like anisotropy favors the $z$ directed N\'eel($z$-N\'eel) order at the zero magnetization, the magnetic field beyond the critical field $H_c$ recovers the critical quantum fluctuation(see Fig.\ref{fig1})  and then the system is described by the Tomonaga-Luttinger(TL) liquid,\cite{haldane} which is characterized by the power law decay of the correlation functions:
\begin{eqnarray}
\langle S_n^zS_0^z\rangle &=& M_z^2 - \frac{1}{4\pi^2\eta n^2} +A_1 \frac{\cos2k_F n}{|n|^{1/\eta}}\cdots \label{lgcf}\\
\langle S_n^xS_0^x\rangle &=& (-1)^n\left[B_0 \frac{1}{|n|^\eta} - B_1 \frac{\cos 2k_F n}{|n|^{\eta+1/\eta}}\cdots \right]
\label{tvcf}\end{eqnarray}
where $\eta$ is the TL exponent, $M_z$ is the uniform magnetization due to a magnetic field $H$,  and  the corresponding Fermi wave number is $ k_F \equiv \pi(1/2-M_z) $. 
The nonuniversal coefficients $A$ and $B$ were evaluated in Ref.\cite{HF}. 
For the isotropic Heisenberg model, $\eta <1$ is always satisfied and thus the transverse fluctuation of (\ref{tvcf}) is dominant. 
For the Ising-like case, however,  $\eta >1$ appears in the low field region, where the longitudinal IC fluctuation becomes dominant. In the actual quasi-1D compound, there is indispensable inter-chain interaction, which may bring the finite temperature phase transition accompanying the field-dependent IC order. 

In this paper, we investigate the field induced IC order for the coupled XXZ chains, using the bosonization combined with the mean-field treatment of the inter-chain interaction\cite{schulz1}.
In particular, we make quantitative analysis of the transition temperatures, taking account of the non-universal coefficients in (\ref{lgcf}) and (\ref{tvcf}).
We then find that the  IC order is certainly realized in the low field region, in contrast with Ref.\cite{wessel} where the possibility of the IC order was not taken into account, while the transverse staggered order appears in the high-field region.
Moreover, we show that the present theory successfully explains the field dependence of the experimentally observed transition temperature.
We also determine the inter-chain coupling of the BaCo$_2$V$_2$O$_8$ as $0.09$K.

This paper is organized as follows.
In Sec. II, we explain the model and the mean-field theory for the inter-chain interaction on the basis of bosonization. 
In Sec.III,  magnetic-field dependences of the transition temperatures are presented for the IC order and the transverse staggered order.  Then the IC order of BaCo$_2$V$_2$O$_8$ is discussed in detail. 
In Sec. IV, we summarize conclusions and discuss related topics.

\section{model and formulation}

The relevant model we consider here is the weakly coupled $S=1/2$ XXZ chains on the simple cubic lattice, whose Hamiltonian is given by
\begin{eqnarray}
{\cal H} &=& \sum_{n,j}[J (\vf{S}_{n,j}\cdot\vf{S}_{n+1,j})_\Delta\nonumber\\
&&+\sum_{n,\langle j,j'\rangle}J' (\vf{S}_{n,j}\cdot\vf{S}_{n,j'})_\Delta- H \sum_{n,j} S_{n,j}^z,
\label{xxz}
\end{eqnarray}
where  $(\vf{S}\cdot\vf{S})_\Delta\equiv S^xS^x+S^yS^y+\Delta S^zS^z$ is the deformed inner product.
The index $n$ runs along the chain direction, $j$ labels the inter-chain directions, and $\langle j,j'\rangle$ denotes the nearest neighbor pair of the chains. 
Then $J$ is the exchange coupling along the chain direction and the inter-chain interaction is controlled by $J'(\ll J)$. Note that we set the lattice space to be unity.

\begin{figure}[t]
\epsfig{file=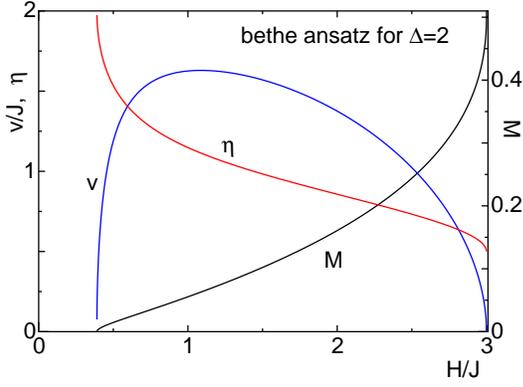,width=7cm}
\caption{(online color) The exact magnetization curve $M$, critical exponent $\eta$ and spin wave velocity $v$ for the XXZ chain of $\Delta=2$.} 
\label{fig1}
\end{figure}

Let us discuss the order-disorder transition in $H>H_c$. 
Since the simple cubic lattice is considered, we can set up the ``sub-chain''  mean fields as
$\vf{S}_{n,j}=\vf{M}_n + \delta\vf{S}_{n,j}$, where $\vf{M}_n$ is a classical vector field.
We assume the IC oscillation of the magnetization of the $z$ component around the uniform(average) magnetization, 
\begin{eqnarray}
 \vf{M}_n& =& [M_z \pm m_{ic}\cos(2k_F n)] \vf{e}_z, 
\end{eqnarray}
where $m_{ic}$ is the amplitude of the IC fluctuation around the average magnetization $M_z$ and the sign depends on the sub chain.
Also for the transverse staggered fluctuation, we assume
\begin{eqnarray}
 \vf{M}_n& =& M_z \vf{e}_z \pm (-)^n m_s \vf{e}_x      
\end{eqnarray}
where $m_{s}$ is the amplitude of the staggered magnetization in the transverse direction.
Then, the mean-field treatment of the inter-chain coupling yields the mean-field Hamiltonian per chain as 
\begin{eqnarray}
{\cal H}_{\rm MF} = {\cal H}_{\rm 1D} + {\cal H}'_{ic/s} \label{MFH}
\end{eqnarray}
where 
\begin{equation}
 {\cal H}_{\rm 1D}\equiv \sum_{n}J (\vf{S}_{n}\cdot\vf{S}_{n+1})_\Delta - \bar{H} \sum_{n} S_{n}^z
\label{1DH}
\end{equation}
is the 1D XXZ chain in an effective magnetic field $\bar{H}$ and ${\cal H}'_{ic/s}$ is the perturbation originating from the mean fields.
For the IC order, we have $\bar{H}\equiv H-z\Delta J' M_z$ and
\begin{eqnarray}
{\cal H}'_{ic}=  - h_{ic}\sum_n\cos(2k_F n) S^z_n + {\rm const}. \label{pszinc}
\end{eqnarray}
where  $ h_{ic}\equiv z \Delta J' m $.
Note that the coordination number is $z=4$ for 3D.
These effective fields should be determined self-consistently.
Also for the transverse staggered fluctuation, we have $\bar{H}\equiv H-z\Delta  J'M_z$ and
\begin{equation}
{\cal H}_s= - h_s\sum_n (-)^n S^x_n + {\rm const} \label{psxst}
\end{equation}
with  $ h_s \equiv z J' m_s$

In order to treat the IC nature in the mean-field Hamiltonian, it is useful to 
 employ the effective model in the continuum limit. This is achieved by the standard bosonization scheme.\cite{giamarchi}
Assuming $\bar{H}-H_c \gg h_{ic}$ or $h_s$,  we can write the XXZ chain in the magnetic field as
\begin{eqnarray}
{\cal H}_{\rm 1D}  \to  \frac{v}{2}\int dx [ (\partial_x \phi)^2 + (\partial_x \theta)^2 ],
\label{TL}
\end{eqnarray}
where $v$ is the spin wave velocity and the compactification radius $R$ is defined by the boundary condition $\displaystyle \phi(x+L) = \phi(x) + \pi R N$.
The equal time commutator of the bosonic fields is defined as $[\phi(x),\theta(y)]=i\Theta(x-y)$, where $\Theta(x-y)$ is the step function.
For the XXZ model, the radius varies $ R=1/\sqrt{\pi} \to 1/\sqrt{4\pi}$ as the magnetic field increasing from $H_c$ to the saturation(free fermion) limit.
The TL exponent is given by $\eta=2\pi R^2$.

The boson representation of the spin operators is given by the formula
\begin{eqnarray}
 S^z_n &\simeq &M_z + \frac{\partial_x \phi(n)}{2\pi R} + a \cos(\frac{\phi(n)}{R}-2k_Fn) \label{bsz}\\
S^+_n &\simeq& (-)^n b e^{i2\pi R\theta(n)}\label{bsp}
\end{eqnarray}
where $a$ and $b$ are nonuniversal coefficients depending on $\Delta$ and $H$. 
Using (\ref{bsz}) and (\ref{bsp}), the amplitude of the equal time correlation function is given by $A_1=a^2/2$, $B_0=b^2/2$ with the regularization
\begin{equation}
\int_0^\infty dk \frac{e^{-\alpha k}}{k}(1-\cos kx) =\ln x \label{regularization},
\end{equation}
where $\alpha$ is a cutoff parameter.
Although the analytical expression of these coefficients in the magnetic field are still unknown, the numerical value is available in Ref.\cite{HF}, which play a crucial role to semi-quantitative evaluation of the transition temperature, as will be seen later.
Substituting (\ref{bsz}) and (\ref{bsp}) into (\ref{pszinc}) and (\ref{psxst}), we obtain the boson field representation of the perturbations,
\begin{eqnarray}
 {\cal H}'_{ic}& \to& -a h_{ic}\int dx  \cos(\frac{\phi}{R})\label{pinc}\\
 {\cal H}'_{s} &\to&  -b h_s \int dx \cos(2\pi R \theta ) \label{pstag}
\end{eqnarray}
where we have omitted the $2k_F$ and $4k_F$ oscillating terms.

Let us consider the effects of ${\cal H}_{ic/s}$ on  the Hamiltonian (\ref{TL}).
The gap generated by the staggered field (\ref{pstag}) was analyzed in Ref.\cite{OA} to be $\Delta E_{s} \sim h_{s}^{2/(4-\eta)}$.
Also for the operator of $\cos(\phi/R)$, the similar standard renormalization group argument leads  $\Delta E_{ic} \sim h_{ic}^{2/(4-1/\eta)}$.
Both of the operators are always relevant between $H_c$ and saturation field.
In the lower field region,  however, $\eta > 1$ and  thus $2/(4-1/\eta)  < 2/(4-\eta)$;
The IC field is more relevant in the low field region, while the transverse staggered field is more relevant in the high field region. 
The border is just $\eta=1$ namely, an effective SU(2) point.
For the case of $\Delta=2$, it corresponds to $H/J \simeq 1.5$ as in Fig.\ref{fig1}.  
This analysis of the gap is consistent with the naive expectation of the critical exponent $\eta$,
which leads the IC order in the low field region.
For the quantitative analysis,  however, the coefficients of the gaps become important.
It should be also noted that the assumption $\bar{H}-H_c \gg h_{ic},\; h_s$ is not valid in the vicinity of the lower critical field.

Now we proceed finite temperature behaviors. 
In the framework of the mean-field theory, $h_{ic}$ or $h_s$ should be determined self-consistently.
Write the IC magnetization of (\ref{MFH}) at $T$, $\bar{H}$ and $h_{ic}$ as $f_{ic}(T, \bar{H},h_s)$, and the staggered magnetization at $T$,  $\bar{H}$ and $h_s$ as $f_s(T,\bar{H},h_s)$. The self-consistent equations are written as
$m_{ic} = f_{ic}(T,\bar{H},h_{ic})$ and $m_s = f_s(T,\bar{H},h_s)$, 
combined respectively with $h_{ic}=z \Delta J'm_{ic}$ and $h_s=zJ'm_s$.
Taking $h_{ic}, h_s \to 0$ limits, we can determine the transition temperatures
\begin{equation}
\frac{1}{z J'\Delta}=\chi_{ic},\qquad 
\frac{1}{z J'}=\chi_s 
\label{mftc}
\end{equation}
where
$\chi_{ic}\equiv\partial f_{ic}/\partial h_{ic}|_{h_{ic}=0}$
and $\chi_{s}\equiv \partial f_s/\partial h_s |_{h_s=0}$. 
According to the linear response theory, the dynamical susceptibility
 can be represented  through the correlation function: $\chi_{\alpha\beta}(q,\omega;T) =-i \sum_n \int dt  e^{i\omega t -iqn}\Theta(t) \langle[S^\alpha(n,t),S^\beta(0,0)]\rangle_T$,  where $\langle \cdots \rangle_T$ denotes the average at a temperature $T$. 
 For the TL Hamiltonian (\ref{TL}), this dynamical susceptibility was actually calculated in Ref.\cite{schulz2,giamarchi}.
For the estimation of the transition temperature, the susceptibility at the soft mode is essential:  for the IC order, $\omega=0$ and $q=2k_F$, and for the staggered order, $\omega=0$ and $q=\pi$.
Taking account of (\ref{regularization}), the explicit form of the leading term becomes
\begin{eqnarray}
&\chi_{ic}&= \chi_{zz}(2k_F, 0;T)= \nonumber\\
&& \frac{A_1}{v}\sin\left(\frac{\pi}{2\eta}\right) \left(\frac{2\pi T}{v}\right)^{1/\eta-2}B(\frac{1}{4\eta},1-\frac{1}{2\eta})^2 \label{chiic}\\
&\chi_s&=\chi_{xx}(\pi,0;T)=\nonumber\\ &&\frac{B_0}{v}\sin\left(\frac{\pi\eta}{2}\right)\left(\frac{2\pi T}{v}\right)^{\eta-2} B(\frac{\eta}{4},1-\frac{\eta}{2})^2 \label{chis}
\end{eqnarray}
where $B(x,y)$ is the Euler's beta function.
Note that the cutoff parameter in (\ref{chiic}) and (\ref{chis}) formally corresponds to $\alpha=1$, due to the regularization (\ref{regularization}).
The magnetic-field dependence is implicitly included in $v$, $\eta$, $A_1$ and $B_0$.
Substituting the above susceptibilities into (\ref{mftc}), we obtain the final result of the transition temperatures
\begin{eqnarray}
T_c^{(ic)}&=&\frac{v}{2\pi}\left(z \Delta J' A_1 \frac{\sin\frac{\pi}{2\eta}}{v}B(\frac{1}{4\eta},1-\frac{1}{2\eta})^2 \right)^\frac{\eta}{2\eta-1}\label{tcinc}\\
T_c^{(s)}&=&\frac{v}{2\pi}\left(z J' B_0 \frac{\sin\frac{\pi\eta}{2}}{v}B(\frac{\eta}{4},1-\frac{\eta}{2})^2 \right)^\frac{1}{2-\eta}\label{tcstag}
\end{eqnarray}
In the above expression, $\eta$ and $v$ can be exactly calculated by solving the Bethe ansatz integral equation  as in Fig. \ref{fig1}.
In addition, the non universal amplitudes $A_1$ and $B_0$ can be obtained by using density matrix renormalization group and the bosonization expression of the correlation function for the open boundary system\cite{HF}.
We can thus  calculate $T_c$ semi-quantitatively without any additional parameter.
Here it should be noted that the previous estimation for  $T_c^{(s)}$ was based on the correlation amplitude at the zero magnetic field for $\Delta\le 1$\cite{wessel}.

\section{results}

\subsection{phase diagram}

\begin{figure}[hb]
\epsfig{file=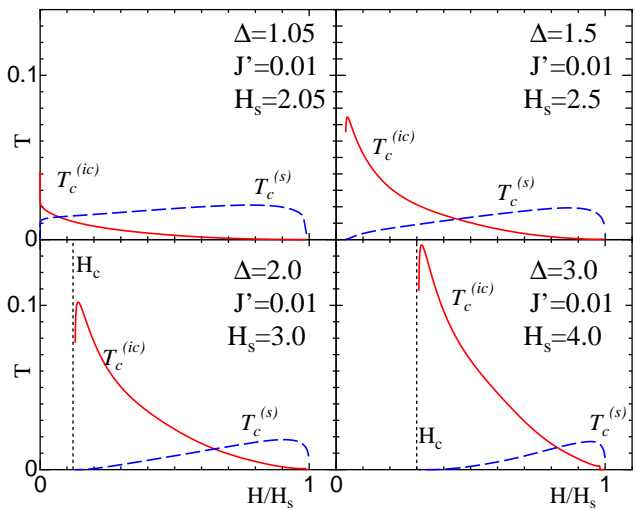,width=8.5cm}
\caption{(online color)The transition temperatures for $\Delta=1.05$, 1.5, 2.0 and 3.0 with $J'=0.01$. 
The intra-chain coupling $J$ is set to be unity.
The solid line and the broken line respectively represent $T_c^{(ic)}$ and $T_c^{(s)}$. 
The vertical dotted lines for $\Delta=2.0$ and 3.0 indicate the critical field $H_c$. 
$H_c$ for $\Delta=1.05$ and 1.5 is not shown here, since it is located in vicinity of $H=0$.
The horizontal axis is normalized by the saturation field $H_s$.}
\label{fig2}
\end{figure}

On the basis of (\ref{tcinc}) and (\ref{tcstag}), we calculate the magnetic field dependences of the transition temperatures of $J'=0.01$ for various $\Delta$.
The correlation amplitudes are extracted from the correlation functions obtained via density matrix renormalization group, as mentioned in the previous section.
In figure 2,  we show the resulting phase diagrams in the $T-H$ plane,  where the solid and broken lines respectively indicate (\ref{tcinc}) and (\ref{tcstag}), and the magnetic field is normalized by the saturation field $H_s$.
The curve corresponding to the higher $T_c$ is realized as an actual order-disorder transition.
In the following, we concentrate on the order-disorder transitions between the critical field $H_c$ and the saturation field $H_s$.
Thus the $z$-N\'eel phase of $M=0$ below $H_c$ is not shown here explicitly. 
In addition, note that $H_c$ for $\Delta=1.05$ and 1.5 is in vicinity of $H=0$ in the scale of Fig.\ref{fig2}.

In Fig.\ref{fig2}, we can see that the IC order certainly occurs above the critical field $H_c$. 
For $\Delta=1.05$, the IC order appears in the vicinity of $H=H_c\simeq 0$.
As $\Delta$ increases, the transverse staggered order is suppressed, while the IC order develops rapidly and the corresponding range of $H$ extends to the higher field region.
An important feature of the IC order is that the field-dependence of $T_c^{(ic)}$ illustrates a characteristic curve; $T_c^{(ic)}$  has the maximum near $H_c$, and it decreases rapidly as $H$ increases.
Such shape of the phase boundary should be contrasted to the semicircle-like boundary for the field-induced staggered-order in the coupled Haldane system.
As further increasing $H$,  the curves for $T_c^{(ic)}$ and  $T_c^{(s)}$ intersect at a certain magnetic field, which is denoted as $H^*$ henceforth, and the transverse staggered order appears for $H_s>H>H^*$.
We can also see that, as $\Delta$ becomes large, $H^*$  shifts to the higher field side.

The behaviors above are basically consistent with the argument based on the TL exponents of the XXZ chain, since the region of $\eta>1$ appears above the critical field $H_c$ and it extends rapidly to the higher field side,  as $\Delta$ is increased.
However, it should be noted that $H^*$ does not coincide with the effective SU(2) point $\eta=1$.
This is because $\eta=1$ is achieved  in the effective field theory level and thus  $A_1\ne B_0$ is permitted even at $\eta=1$, in contrast to the isotropic Heisenberg chain at the zero field having the SU(2) symmetry in the spin operator level. 
Since the correlation amplitude $A_1$ has a larger value than $B_0$(e.g. see FIG. 2 in Ref.\cite{HF}), $T_c^{(ic)}$ is relatively enhanced than $T_c^{(s)}$, implying that  $H^*$ slightly shifts to the higher field side than the effective SU(2) point. 
In this sense, the precise amplitudes are essential in the inter-chain mean-field theory.
In addition, we can see that  $\Delta$ in (\ref{tcinc}) is also a source of such an  enhancement of the IC order.

We next discuss the inter-chain-coupling dependence of transition temperatures.
According to eqs. (\ref{tcinc}) and (\ref{tcstag}), the precise $J'$-dependences are given by $T_c^{(ic)}\propto J'^{\eta/(2\eta-1)}$ and $T_c^{(s)}\propto J'^{1/(2-\eta)}$, so that  the scale of the transition temperature naturally becomes large, as $J'$ is increased.
In particular, we can see that $T_c^{(ic)}$ is more easily lifted toward the higher field region where $\eta<1$,  since, as mentioned above, the IC order is basically enhanced by the correlation amplitude $A_1$ and the anisotropy $\Delta$.
However, we should remark that such enhancement of  $T_c^{(ic)}$ in the inter-chain mean-field theory does not always lead to a clear observation of the IC order for a larger $J'$; we need to pay special attention to stability of the IC order.
Let us recall that the spin flop transition occurs for the case of the spatially isotropic exchange coupling($J=J'$), where the magnetization directly jumps from $z$-Neel phase of $M=0$ to the transverse staggered ordered state.
This suggests that the IC order becomes thermodynamically unstable beyond  a certain critical $J'$,\cite{critical} so that it is embedded in the magnetization jump.
Unfortunately, the critical $J'$ can not determined within the frame work of the mean-field theory for the inter-chain coupling, since analytical calculation of the free energy is still a difficult task. 
In the next subsection, nevertheless, we shall show that the stable IC order actually occurs in the experimental situation of BaCo$_2$V$_2$O$_8$.

\subsection{comparison with experiment}

\begin{figure}[hbt]
\epsfig{file=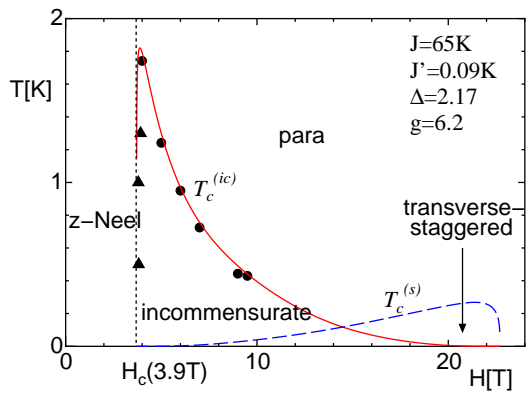,width=7cm}
\caption{(online color)The transition temperatures for the IC order and the transverse staggered order.
The solid line and the broken line respectively represent $T_c^{(ic)}$ and $T_c^{(s)}$ for $\Delta=2.17$ and $J'/J=0.00138$.
The vertical dotted line means $H_c=3.9$T. 
The solid circles indicate the experimentally observed transition temperature\cite{kimura2} and the solid triangles mean the phase boundary between the $z$-N\'eel order and the field induced IC order.
} 
\label{fig3}
\end{figure}

Let us discuss the field dependence of the transition temperature of BaCo$_2$V$_2$O$_8$.
The basic parameters were determined by the magnetization and ESR measurements\cite{kimura1}.
The exchange coupling in the chain direction is given by $J\simeq 65$K and the precise anisotropy parameter is $\Delta\simeq 2.17$. 
The critical field is $H_c\simeq 3.9$T and the saturation field is about 23T with $g=6.2$.
In addition, we have actually calculated the correlation amplitudes of $\Delta=2.17$.
In Figure\ref{fig3}, the field dependence of the transition temperature for  $J'\simeq 0.09$K($J'/J=0.00138$) is illustrated  together with the experimental data,
where the solid and broken lines indicate (\ref{tcinc}) and (\ref{tcstag}) respectively.

In the left side of $H_c$ corresponding to the solid triangles, the $z$-N\'eel order occurs, at which the uniform magnetization is zero.
Note that the transition temperature to the $z$-N\'eel phase at $H=0$ is about 5.4K, which is much higher than $T_c^{(ic)}$.
Above $H_c$, the $z$-N\'eel order is destroyed by the magnetic field and then we come into the targeted region of the present theory.
The solid circles indicate the experimental transition temperature up to 12T.
We can see that the theoretical curve (\ref{tcinc}) excellently reproduces the experimental results, implying the inter-chain mean-field theory is basically correct for 3D.
A remarkable point is that the shape of the experimental phase boundary is consistent with the theoretical curve of the IC order;
As $H$ increases above $H_c$, $T_c^{(ic)}$ decreases rapidly from $T\simeq 1.7$K down to $0.4$K.
The above facts support that the IC order driven by the one dimensionality can be thermodynamically stabilized in the experimental situation.
Another interesting point is that, as further increasing $H$,  the theoretical curves for $T_c^{(ic)}$ and  $T_c^{(s)}$ intersects at $H^*\simeq 15.1$T,  which predicts that the transverse staggered order appears above $H\simeq 15.1$T.
In order to verify the theory,  a specific heat measurement in the higher field is is highly desirable.
However, the value of $T_c^{(s)}$ is relatively low and thus the experimental observation in the competing region may be subtle.

\section{summary and discussions}

We have discussed the field induced IC order on the basis of the bosonization combined with the mean-field theory for the inter-chain interaction.
In particular, the numerically exact correlation amplitudes plays the crucial role to explain the shape of the experimental phase boundary.
In order to investigate the IC order beyond the mean-field level, we have also performed quantum montecarlo (QMC) simulations based on the directed loop algorithm. 
Then, we have confirmed that the IC order actually occurs for a $J'/J=0.1$\cite{suzuki-kawashima}.
We can therefore conclude that the field-induced IC order is certainly realized in the actual system and the inter-chain mean-field treatment captures the essential nature of it.
The inter-chain coupling of BaCo$_2$V$_2$O$_8$ estimated within the mean-field theory is  $J'\simeq 0.09$K.

From theoretical point of view, the phase transition for the 3D classical spin model with the easy-axis anisotropy was intensively studied in 70s, in the context of the spin flop transition.\cite{fisher}
The low magnetized state is unstable in the 3D isotropic lattice system and the magnetization jumps directly from the $z$-N\'eel state into the spin flopped state.
The spin flop transition also occurs for the 2D Ising-like XXZ model on the isotropic square lattice at the zero temperature.\cite{kohno-takahashi}
The present result implies that, as the 1D fluctuation is enhanced,  the IC order ---spin version of the charge density wave(CDW)--- emerges in the phase diagram.
Of course, BaCo$_2$V$_2$O$_8$ is insulating, and thus the mechanism is attributed to the nesting of ``spin'' itself.
In this sense, the present IC order is very similar to that in the spin-Peierls system.\cite{cross} 
However, the driving mechanism is the inter-chain spin-spin interaction itself rather than a spin-phonon coupling in the spin-Peierls case.
Since the inter-chain interaction favors the transverse staggered order as well, the spin flop transition may be induced with a certain finite inter-chain coupling, implying that the thermodynamic stability of the IC order is a non-trivial question.
The present result demonstrate that the IC order based on the 1D mechanism is certainly stabilized in the actual experimental situation.
For the quasi-1D spin model, the Fermi wave number $k_F$  can be easily  controlled by the magnetic field, in contrast with the CDW in the metallic system. 
A further experimental study, particularly neutron scattering experiment in the magnetic field, is highly interesting.
In addition, the connection to the spin flop transition in the high field region is also theoretically important problem,  although the experiment for BaCo$_2$V$_2$O$_8$ suggests a weak first order transition at $H_c$ accompanying the spin-lattice coupling which may cooperatively stabilize the incommensurate order.

Finally we remark that our theory is valid not only for the similar quasi-1D systems with the easy axis anisotropy, but also for a class of the frustrating systems.
In fact, the frustrating systems is often mapped into an effective XXZ model, for which the IC order is actually pointed out.\cite{maeshima,suzuki-suga}
Such enhancement of the IC fluctuation is also reported for an anisotropic $S=1$ chain\cite{sakai}.
We hope that the rich physics associated with the spin anisotropy and quantum fluctuation can be developed by further theoretical and experimental researches.

\acknowledgments

We would like to thank S. Kimura, M. Sato and  N. Kawashima for valuable discussion.
We are also grateful T. Hikihara for providing the numerical data in Ref.\cite{HF}.
This work was partially supported by Grants in Aid for Science Researches from MEXT, Japan.
It was also supported by ``High Field Spin Science in 100T"


\begin{thebibliography}{99}
\bibitem{critical} This critical $J'$ may depend on the magnetic field $H$.
\end{thebibliography}
\end{document}